\documentclass[conference]{IEEEtran}
\IEEEoverridecommandlockouts
\usepackage{cite}
\usepackage{amsmath,amssymb,amsfonts}
\usepackage{algorithmic}
\usepackage{graphicx}
\usepackage{textcomp}
\usepackage{xcolor}
\usepackage[a4paper, total={184mm,239mm}]{geometry}
\def\BibTeX{{\rm B\kern-.05em{\sc i\kern-.025em b}\kern-.08em
    T\kern-.1667em\lower.7ex\hbox{E}\kern-.125emX}}

\usepackage{url}
\usepackage{bm}
\usepackage{multirow}

\graphicspath{{figs_sesr/}}

\begin{document}

\title{Super-Efficient Super Resolution for Fast Adversarial Defense at the Edge\thanks{This preprint is for personal use only. The official article will appear in proceedings of Design, Automation \& Test in Europe (DATE), 2022, as part of the Special Initiative on Autonomous Systems Design (ASD).}}

\author{\IEEEauthorblockN{
Kartikeya Bhardwaj$^1$, Dibakar Gope$^2$, James Ward$^3$,  Paul Whatmough$^4$, and Danny Loh$^1$}
\IEEEauthorblockA{$^1$\textit{Arm Inc., San Jose, CA, USA}, $^2$\textit{Arm Research, Austin, TX, USA} \\
$^3$\textit{Arm Inc., Galway, Ireland}, $^4$\textit{Arm Research, Boston, MA, USA}\\
\{kartikeya.bhardwaj, dibakar.gope, james.ward, paul.whatmough, danny.loh\}@arm.com}
}

\maketitle

\begin{abstract}
Autonomous systems are highly vulnerable to a variety of adversarial attacks on Deep Neural Networks (DNNs). Training-free model-agnostic defenses have recently gained popularity due to their speed, ease of deployment, and ability to work across many DNNs. To this end, a new technique has emerged for mitigating attacks on image classification DNNs, namely, preprocessing adversarial images using super resolution -- upscaling low-quality inputs into high-resolution images. This defense requires running both image classifiers and super resolution models on constrained autonomous systems. However, super resolution incurs a heavy computational cost. Therefore, in this paper, we investigate the following question: Does the robustness of image classifiers suffer if we use tiny super resolution models? To answer this, we first review a recent work called Super-Efficient Super Resolution (SESR)~\cite{sesr} that achieves similar or better image quality than prior art while requiring $\bm{2\times}$ to $\bm{330\times}$ fewer Multiply-Accumulate (MAC) operations. We demonstrate that despite being orders of magnitude smaller than existing models, SESR achieves the same level of robustness as significantly larger networks. Finally, we estimate end-to-end performance of super resolution-based defenses on a commercial Arm Ethos-U55 micro-NPU. Our findings show that SESR achieves nearly $\bm{3\times}$ higher FPS than a baseline while achieving similar robustness.
\end{abstract}


\section{Introduction}\label{sec:intro}
With the rise of autonomous systems, there is now an incredible demand for robust deep learning at the edge. Current autonomous systems are highly resource-constrained and need to maintain high performance under strict energy consumption constraints. To improve robustness, many state-of-the-art defense techniques rely on adversarial training (or robust training) which comes at a heavy computational cost. Specifically, adversarial training can take up to ten times the normal forward and backward pass steps compared to a conventional training method~\cite{hua2021bullettrain}. Furthermore, adversarial training may not even be possible in a practical system if the model is being deployed on a given hardware by a third party and they cannot perform any training (say, due to a lack of access to the third party model's internal parameters/structure or datasets/adversarial samples due to privacy or regulatory concerns or even lack of expertise)~\cite{tip2019, bhardwaj2019dream, nagel2019data}. Hence, \textit{training-free} adversarial defense mechanisms have gained equal importance in the recent years~\cite{xie2018mitigating, prakash2018deflecting, tip2019, das2017keeping, cheng2021defense}. These techniques are particularly important in autonomous systems that deploy a large number of DNNs and it may not be possible to defend all of them effectively via robust training. Therefore, a model-agnostic, fast, training-free defense of modern inference pipelines is highly important. 

For this model-agnostic training-free adversarial defense, it has been recently proposed that image Super Resolution (SR) can significantly improve the robustness of deep networks~\cite{tip2019}. Specifically, Mustafa \textit{et al.}~\cite{tip2019} provide a comprehensive study where pretrained models such as Enhanced Deep Super Resolution (EDSR) networks~\cite{edsr} and denoising methods such as wavelet denoising are used as preprocessing steps to defend against adversarial attacks. The main idea behind this approach is that since SR models are trained to output images on the natural image manifold, they can bring off-the-manifold images (e.g., those perturbed due to attacks) back to the natural image manifold by introducing high-frequency feature details into the image. Note that, this training-free model-agnostic defense works in the gray-box settings~\cite{tip2019, guo2018countering}, where the classification model under attack is fully known to the adversary, but the defense method is unknown. We will assume the same gray-box attack settings in this work. Practically, this is a reasonable assumption since even if attackers have access to the classification models, the SR network may not be accessible to them (particularly if the models are being deployed on a specific hardware by a third party). 

To this end, it has been shown that models like EDSR are highly successful at this defense~\cite{tip2019}. However, with nearly 42M parameters, EDSR-like methods are completely infeasible for constrained autonomous systems. Hence, in this paper, we address the following \textbf{key question}: \textit{Does the robustness of SR-based training-free defense methods suffer if we use extremely tiny SR models that are suitable for deployment on resource-constrained devices?} To answer this, we first review a recent network called Super-Efficient Super Resolution (SESR)~\cite{sesr} that achieves state-of-the-art results in hardware-efficient SR. SESR has been demonstrated to achieve similar or better image quality (in terms of PSNR) than existing SR methods while using $2\times$-$330\times$ fewer Multiply-Accumulate (MAC) operations. We then explore how the robustness of classification methods changes as a function of the computational complexity of SR models.  
Overall, we make the following \textbf{key contributions}:
\begin{itemize}
    \item We discover that tiny SR networks such as SESR~\cite{sesr} or other tiny publicly available networks such as FSRCNN~\cite{fsrcnn} can result in nearly as much robustness as large networks like EDSR-base or EDSR~\cite{edsr} against a variety of adversarial attacks. SESR models are two orders of magnitude smaller than methods like EDSR-base, thus making training-free adversarial defense mechanisms feasible for highly resource constrained devices. We show improvements of up to $6\times$ in MACs over FSRCNN with similar robustness accuracy.
    \item We demonstrate concrete latency improvements for SESR-based training-free defense by using performance estimators for commercial micro-Neural Processing Units (micro-NPUs, i.e., microcontroller-scale AI accelerators) such as Arm Ethos-U55 which can accompany Arm Cortex-M (microcontroller-based) systems. Our results show that end-to-end latency (i.e., combined classification and SR) for SESR improves by up to $3\times$ compared to FSRCNN on Ethos-U55.
\end{itemize}

The rest of the paper is organized as follows: Section~\ref{sec:rel} describes the related work. Next, Section~\ref{sec:app} first explains the SR-based training-free adversarial defense pipeline for DNN inference and then reviews the SESR model structure and its training methodology. Extensive empirical results are then presented in Section~\ref{sec:exp}. Finally, after some discussion on open problems in Section~\ref{sec:dis}, the paper is concluded in Section~\ref{sec:conc}.

\section{Related Work}\label{sec:rel}
One of the most successful defense methods is adversarial or robust training, which, unlike other defense measures, tries to improve the fundamental robustness of deep networks. Adversarial training does this by supplementing training data with adversarial instances in each training loop. As a result, when confronted with attacked examples, adversarially-trained models are more robust than conventional models. 
In recent years, numerous research efforts have been devoted to exploring mechanisms that directly or implicitly address robustness in deep learning by either adding adversarial
examples to the dataset or incorporating them into the objective function for optimization~\cite{Wang2020Improving, Wong2020Fast}.
The prominent attacks on machine learning models that we focus on in this work rely on the gradient of the model to estimate the local optima perturbations that will deceive the classifier. In addition to adversarial training, a body of research focusing on robust optimization investigates the use of regularization algorithms to limit the impact of tiny perturbations in the input on output decisions.
This stream of works~\cite{xie2019denoising, zhangicml2019} applies a regularization penalty to the functional loss in order to lessen the network's sensitivity to small perturbations in the inputs.

The above model-specific techniques frequently result in deep learning models with reduced generalization capacity. Despite substantial attempts, robust training has not completely overcome the sensitivity of deep learning models to attacks.
Furthermore, adversarially trained or regularized models that are resistant to a certain attack can be easily defeated by other types of attacks. Hence, the dependability of these model-specific techniques suffers greatly as a result of this type of poor generalization to different attacks. As a result, training-free model-agnostic defense mechanisms that can defend against adversarial attacks without knowing the target model's architecture or parameters have emerged as a  promising approach for real-world deployment.

In comparison to model-specific defense mechanisms, model-agnostic methods use input transformations to remove adversarial perturbations before feeding them into neural network classifiers. A number of input transformation methods, such as image cropping-rescaling, bit-depth reduction, JPEG compression~\cite{das2017keeping}, total variance minimization, and image quilting, or a combination of these methods, have been proposed in the literature to suppress the human-imperceptible, high-frequency noise components and improve the model's robustness. In~\cite{guo2018countering}, a combination of total variation minimization and image quilting is used to defend against strong attacks. \cite{prakash2018deflecting} deflects attention by carefully altering less crucial image pixels, where they randomly sample a pixel from an image, and replace it with another randomly selected pixel from within a small square neighborhood. Denoising in the wavelet domain has been demonstrated to outperform other approaches such as total variation minimization~\cite{tip2019, prakash2018deflecting}. \cite{xie2018mitigating} performs image alterations by padding an image and taking multiple random crops before sending the ensembles through a CNN classifier. 

The biggest obstacle facing most transformation-based defenses is that the transformation degrades the quality of non-adversarial images, leading to a loss of accuracy. This has hampered the effectiveness of transformations as a practical defense, as even those that are good at removing adversarial transformations fail to keep the model accurate on clean images. Although previous image transformation techniques introduced artifacts while countering adversarial noise, the recently proposed SR-based image transformation scheme~\cite{tip2019} demonstrates the potential to preserve critical image content while having a minimal impact on the classifier's performance on clean, non-attacked images. However, the high computational complexity and significant runtime cost of the high-end SR models utilized in \cite{tip2019} makes them unsuitable for defense against adversarial attacks on resource-constrained devices. To this end, our work is the first to show the true capability of tiny SR networks in successfully defending CNN classifiers and maintaining robustness comparable to that of high-end SR models.

\section{Approach}\label{sec:app}
In this section, we first describe the main idea behind SR-based adversarial defense~\cite{tip2019} and the precise problem we address in this paper. Then, we review a recent development in efficient SR methods, namely, the SESR network~\cite{sesr}.
\begin{figure*}[tb]
\centering
\includegraphics[width=1.0\textwidth]{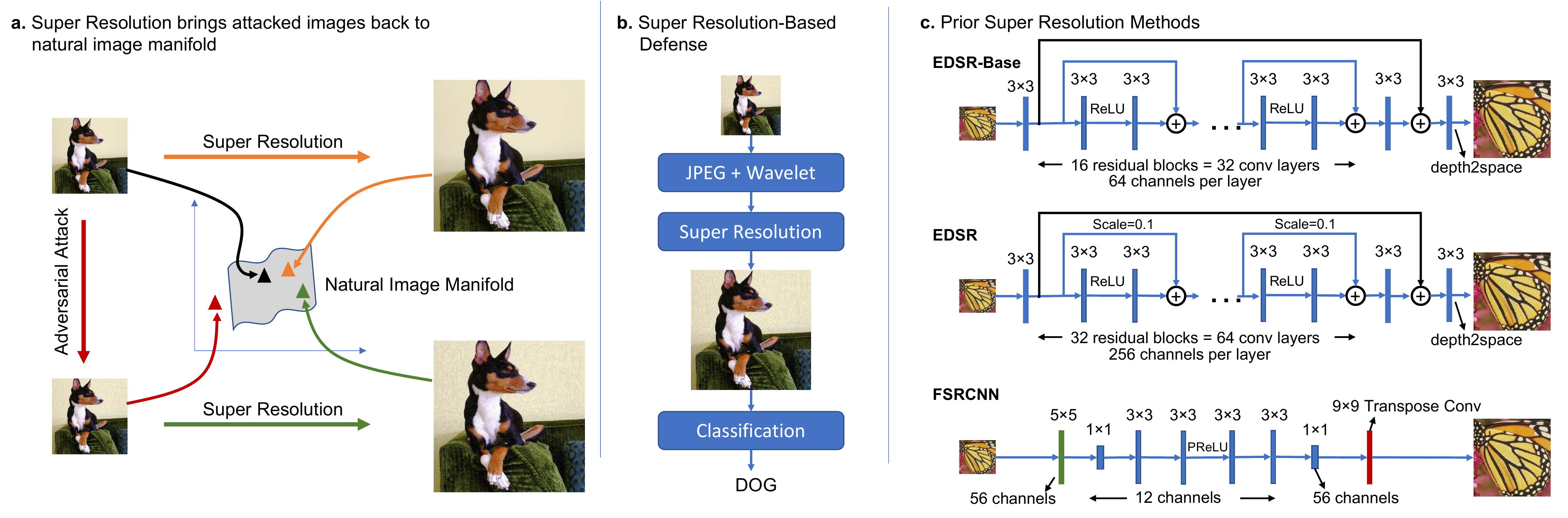}\vspace{-3mm}
	\caption{Overview of the approach: (a)~As explained in~\cite{tip2019}, adversarial attacks move the images out of the natural image manifold. However, performing SR on the attacked images adds important information back into the image, thereby moving it back into the natural image manifold. Therefore, conducting SR on attacked images can improve adversarial robustness in the gray-box settings. (b)~The defense mechanism used in this paper for analysis of various SR networks. (c)~Structure of prior SR methods like EDSR-base~\cite{edsr}, EDSR~\cite{edsr}, and FSRCNN~\cite{fsrcnn}.}
\label{fig:app}
\end{figure*}

\subsection{Adversarial Defense using Super Resolution}\label{sec:appAdvSR}
When adversarial perturbations are added to the images, they are moved out of the natural image manifold~\cite{tip2019}. As a result, such adversarial images are easily misclassified by the downstream classification models. On the other hand, SR networks are trained to take a low-resolution input and generate a high-resolution output image. The typical loss functions used to train SR are Mean Absolute Error (MAE)~\cite{edsr}, Mean Squared Error (MSE)~\cite{fsrcnn}, or even some advanced methods like Generative Adversarial Networks (GANs)~\cite{srgan}. The MAE/MSE loss functions try to minimize the error between the network's output and high-resolution training images. Hence, the SR networks learn to output images on the natural image manifold. Therefore, when adversarial images are passed through a SR model, they are mapped back to the natural image manifold. This is illustrated in Fig.~\ref{fig:app}(a) and, as explored in~\cite{tip2019}, is the main idea behind SR-based defense against gray-box attacks.

Inspired by~\cite{tip2019}, we study the SR-based adversarial defense from a computational cost standpoint which is important to practically deploy this defense mechanism on resource-constrained autonomous systems. Specifically, we show our defense method in Fig.~\ref{fig:app}(b). We start with JPEG compression~\cite{das2017keeping} and wavelet denoising~\cite{prakash2018deflecting, tip2019} of the input images. Then, we perform $\times2$ SR on the denoised, compressed images. Finally, the upscaled images are passed through the classification network. Note that, this is a training-free defense method: Neither the super resolution method nor the classification networks need to be adversarially trained. This is why the defense works for gray-box attacks where the classification network is fully known to the attackers but the SR network is unknown. Therefore, they cannot pass the gradient through the SR model to craft fully-white-box adversarial images. Gray-box attacks are known to be more difficult than black-box attacks~\cite{guo2018countering}.

So far, the prior work does not focus on the computational efficiency of SR-based defense of inference pipelines. Specifically, Mustafa \textit{et al.}~\cite{tip2019} have used large models like EDSR~\cite{edsr} to evaluate robustness. The structure of EDSR-base and the complete EDSR network is shown in Fig.~\ref{fig:app}(c, top, center). As evident, the EDSR network has more than 64 layers, most of them containing 256 channels. This results in nearly 42M parameters. Such SR networks are prohibitively expensive and \textit{cannot} be deployed on resource-constrained edge devices. Therefore, in this paper, we address the following problem: \textit{Does the SR-based adversarial defense remain effective if we use extremely small SR networks that can be deployed on constrained autonomous system hardware?} Hence, we explore the use of tiny SR networks for this gray-box defense. One such tiny SR network called FSRCNN~\cite{fsrcnn} is shown in Fig.~\ref{fig:app}(c). FSRCNN is a small VGG-style network (i.e., no residual connections). Compared to EDSR, FSRCNN only uses 24.3K parameters and is orders of magnitude more efficient. In the next section, we will describe another method called SESR~\cite{sesr} which is even more efficient than FSRCNN.

\subsection{Super-Efficient Super Resolution (SESR)}\label{sec:appSESR}
The SESR network~\cite{sesr} exploits linear overparameterization~\cite{ExpandNets2020, AroraICML2018} to achieve state-of-the-art results in efficient SR. The idea is to train a very large network that can be analytically collapsed into a very efficient inference network. This is possible because SESR relies on \textit{Collapsible Linear Blocks} which first expand the number of channels from $f_i$ to $p$ ($p>>f_i$) using a $k\times k$ convolution, and then project it back to $f_o$ channels using a $1\times 1$ convolution ($p>>f_o$). If the number of input channels ($f_i$) and output channels ($f_o$) are equal, the inputs are added to the output via a short-residual. The key property of a Collapsible Linear Block is that it does not contain any non-linear activation functions. Hence, they can be analytically collapsed into a single $k\times k\times f_i \times f_o$ convolution (i.e., the expansion stage disappears and the resulting layer is highly efficient). Likewise, the short-residual can also be collapsed into the convolution weights (see~\cite{sesr} for details).

\begin{figure}[tb]
\centering
\includegraphics[width=0.45\textwidth]{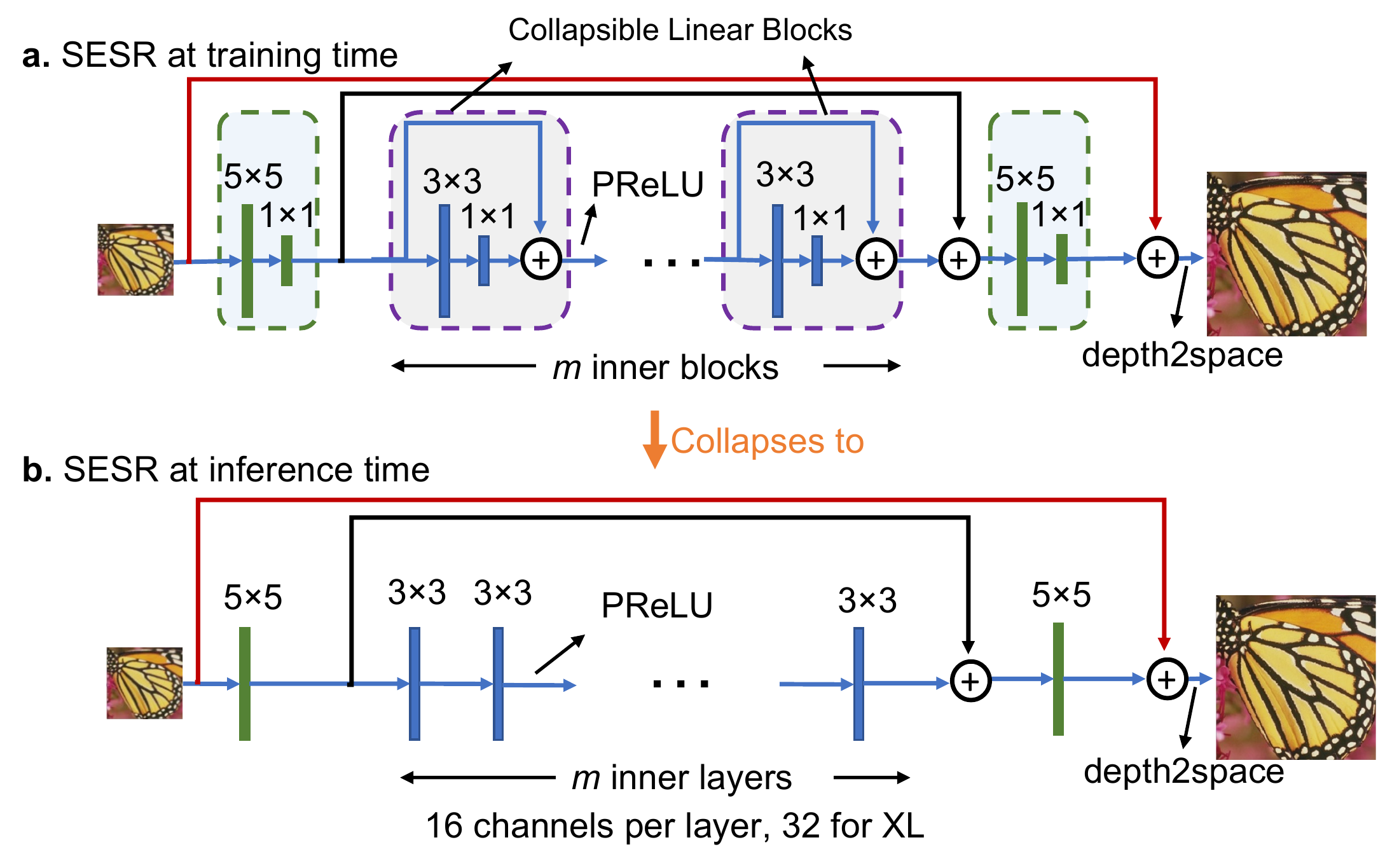}\vspace{-3mm}
	\caption{Super-Efficient Super Resolution (SESR) network~\cite{sesr}. At training time, a large network shown in (a) is trained that can be analytically collapsed into a highly efficient network shown in (b). SESR achieves state-of-the-art results.}
\label{fig:sesr}
\end{figure}

Thus, the large network shown in Fig.~\ref{fig:sesr}(a) collapses into the hardware-efficient network shown in Fig.~\ref{fig:sesr}(b). The inference-time network resembles a VGG-style network like FSRCNN except that it also has two long-residuals. The small SESR networks that contain merely $f_i = f_o = 16$ channels at intermediate layers are named as SESR-M$\{m\}$, where $m\in \{2,3,5\}$ is the number of $3\times 3$ layers in the inference network. Meanwhile, a larger network called SESR-XL contains 32 intermediate channels and eleven $3\times3$ layers. Overall, SESR methods achieve state-of-the-art results in efficient SR with nearly $\bm{2\times}$-$\bm{330\times}$ reduction in MACs over prior art with similar or better PSNR. This allows SESR to achieve \textit{real-time} (up to 36-46FPS) when performing extremely expensive [1080p to 4K] SR task on a tiny \textit{mobile-NPU} (e.g., Arm Ethos-N78 NPU suitable for deployment in mobile devices like phones, TVs, displays, etc.)~\cite{sesr}. In addition to hardware-efficiency, the use of collapsible linear blocks has advantages in deep learning optimization properties (see~\cite{sesr, AroraICML2018} for details). Next, we propose to use these tiny SESR networks and analyze their impact on adversarial robustness in gray-box attack settings. 

\section{Experimental Setup and Results}\label{sec:exp}
\subsection{Experimental Setup}\label{sec:expSetup}
\textbf{Baseline classifiers and datasets. }
We produce adversarial images by attacking three different classifiers: MobileNet-V2, ResNet-50, and Inception-V3. We evaluate and compare different SR-based defense mechanisms for these three classifiers. All experiments are carried out on a subset of $5000$ images from the ImageNet (ILSVRC) validation set. For each classifier, the $5000$ images are chosen such that the corresponding model achieves a top-1 accuracy of $100\%$ on the clean, non-attacked images since it is not useful to evaluate defense mechanisms on images that have already been misclassified.

\textbf{Adversarial attacks. }
Throughout this paper, we use four standard attack techniques: Fast Gradient Sign Method (FGSM)~\cite{fgsm}, Projected Gradient Descent (PGD)~\cite{pgd}, Auto-PGD (APGD)~\cite{apgd}, and Diverse Input Iterative FGSM (DI$^2$FGSM)~\cite{difgsm}. These methods are implemented using publicly available code\footnote{\url{https://github.com/Harry24k/adversarial-attacks-pytorch}} provided by~\cite{kim2020torchattacks}. We use $\epsilon= 8/255$ for all attacks. Adversarial images are passed through the defense method shown in Fig.~\ref{fig:app}(b) within a gray-box setting. 

\textbf{Defense mechanisms. }
We compare our proposed SESR-based defense technique against a number of image upscaling-based methods. Specifically, we investigate two major types of image upscaling techniques: interpolation-based methods and deep learning-based methods. While interpolation-based methods, such as Nearest Neighbor, have a low computational complexity, they achieve poor image quality. Deep learning-based SR techniques, on the other hand, have been shown to significantly outperform interpolation-based methods in terms of PSNR. We compare the performance of SESR-based defense to that of recently proposed state-of-the-art deep learning-based SR techniques such as EDSR and FSRCNN in protecting against adversarial attacks. Of note, the SR networks upscale the input image from $299\times 299$ to $598\times 598$. The classification models then operate on this large $598\times598$ image.

\subsection{PSNR Results for SR methods}\label{sec:expPSNR}
We start by training the various SR networks on the DIV2K dataset with $\times2$ scale. Note that, since classification methods require input in RGB colorspace, we trained our SESR networks as well as EDSR and FSRCNN directly in RGB\footnote{MACs and number of parameters for SESR and FSRCNN in our work are different from those reported in~\cite{sesr,fsrcnn}. The reason is that~\cite{sesr,fsrcnn} perform SR on Y (luma)-channel only after converting the RGB image into Y-Cb-Cr colorspace. Throughout this paper, we will work directly with RGB.}. Table~\ref{tab:PS} shows the PSNR results for different methods on the DIV2K validation set. Here, MACs are for upscaling $299\times299$ images to $598\times 598$. Clearly, SESR-M2 outperforms FSRCNN with $6\times$ fewer MACs. For larger networks, SESR-XL achieves nearly 0.5dB less PSNR but with nearly $10\times$ fewer MACs. In the next section, we evaluate how these different compute cost \textit{vs.} image quality tradeoffs affect adversarial robustness.
\begin{table}[]
\caption{PSNR results (in RGB colorspace) for SR methods}
\label{tab:PS}
\begin{center}
\begin{tabular}{|l|c|c|c|}
\hline
Model & Parameters & MACs & PSNR ($\times$2 SR, DIV2K) \\ \hline
FSRCNN~\cite{fsrcnn} & 24.3K & 5.82B & 32.92 \\ \hline 
EDSR-base~\cite{edsr} & 1.19M & 106B & 34.62 \\ \hline 
EDSR~\cite{edsr} & 42M & 3400B & 35.03 \\ \hline\hline
SESR-M2~\cite{sesr} & 10.6K & 0.948B & 33.26 \\ \hline
SESR-M3~\cite{sesr} & 12.9K & 1.154B & 33.44 \\ \hline
SESR-M5~\cite{sesr} & 17.5K & 1.566B & 33.64 \\ \hline
SESR-XL~\cite{sesr} & 113.3K & 10.13B & 34.14 \\ \hline
\end{tabular}
\end{center}
\end{table}

\begin{table*}[]
\caption{Accuracy results for various classification and super resolution networks across different adversarial attacks}
\label{tab:rob}
\begin{center}
\begin{tabular}{|l|l|c|c|c|c|c|c|}
\hline
Classification Network & SR method   & Parameters & MACs & FGSM~\cite{fgsm} & PGD~\cite{pgd} & APGD~\cite{apgd} & DI$^2$FGSM~\cite{difgsm} \\ \hline\hline
\multirow{9}{*}{MobileNet-V2~\cite{mobilenetv2}} & No Defense   &  $-$  & $-$  &   3.42 & 6.01 & 30.8 & 0.02   \\ \cline{2-8}
                    & Nearest Neighbor  &  $-$  &  $-$  & 10.07 & 15.91 & 21.06 & 6.47   \\ \cline{2-8}
                    & EDSR-base~\cite{edsr}  &  1.19M  &  106B  & 17.46 & 33.37 & 41.77 & 13.14   \\ \cline{2-8} 
                   & EDSR~\cite{edsr}  &  42M  & 3400B  &   17.00 & 32.49 & 40.27 & 13.14   \\ \cline{2-8} 
                   & FSRCNN~\cite{fsrcnn}   &  24.336K  & 5.82B  &   \textbf{19.83} & \textbf{35.02} & \textbf{43.98} & 13.66   \\ \cline{2-8} 
                   & SESR-M2~\cite{sesr}  &  \textbf{10.608K}  & \textbf{0.948B}  &   19.61 & 34.72 & 43.84 & 13.8   \\ \cline{2-8} 
                   & SESR-M3~\cite{sesr} &  12.912K  & 1.154B  &   19.33 & 34.54 & 43.44 & \textbf{13.94}   \\ \cline{2-8} 
                   & SESR-M5~\cite{sesr} &  17.520K  & 1.566B  &   19.15 & 34.76 & 43.3 & \textbf{13.94}   \\ \cline{2-8} 
                   & SESR-XL~\cite{sesr} &  113.3K  & 10.13B  &   18.36 & 33.65 & 42.39 & 13.46   \\ \hline\hline
\multirow{9}{*}{ResNet-50~\cite{resnet}} & No Defense   &  $-$  & $-$  & 8.52 & 17.07 & 22.85 & 0.22  \\ \cline{2-8} 
                    & Nearest Neighbor  &  $-$  &  $-$  &   19.96 &	31.48 &	32.65 &	20.68  \\ \cline{2-8}
                    & EDSR-base~\cite{edsr}  &  1.19M  &  106B  &   31.66 & 48.66 & 50.56 & 30.48  \\ \cline{2-8} 
                   & EDSR~\cite{edsr}  &  42M  & 3400B  &   31.06 & 46.43 & 49.08 & 30.5   \\ \cline{2-8} 
                   & FSRCNN~\cite{fsrcnn}   &  24.336K  & 5.82B  &   \textbf{32.65} & \textbf{49.8} & 51.76 & 31.24   \\ \cline{2-8} 
                   & SESR-M2~\cite{sesr}  &  \textbf{10.608K}  & \textbf{0.948B}  &  32.34 & 49.66 & \textbf{51.82} & 31.24   \\ \cline{2-8} 
                   & SESR-M3~\cite{sesr} &  12.912K  & 1.154B  &   31.96 & 49.46 & 51.74 & \textbf{31.38}   \\ \cline{2-8} 
                   & SESR-M5~\cite{sesr} &  17.520K  & 1.566B  &   32.2 & 49.64 & \textbf{51.82} & 31.2   \\ \cline{2-8} 
                   & SESR-XL~\cite{sesr} &  113.3K  & 10.13B  &   31.92 & 48.96 & 51.24 & 30.48   \\ \hline\hline
\multirow{9}{*}{Inception-V3~\cite{inceptionv3}} & No Defense   &  $-$  & $-$  &   25.89 & 10.24 & 11.42 & 0.52   \\ \cline{2-8} 
                    & Nearest Neighbor  &  $-$  &  $-$  &   58.22 &	69.15 &	71.75 &	51.6   \\ \cline{2-8}
                    & EDSR-base~\cite{edsr}  &  1.19M  &  106B  &   60.22 & 69.55 & 72.17 & 54.92   \\ \cline{2-8} 
                   & EDSR~\cite{edsr}  &  42M  & 3400B  &   60.12 & 69.57 & \textbf{72.49} & \textbf{55.38}   \\ \cline{2-8} 
                   & FSRCNN~\cite{fsrcnn}   &  24.336K  & 5.82B  &  60.12 & \textbf{69.93} & 71.97 & 54.24   \\ \cline{2-8} 
                   & SESR-M2~\cite{sesr}  &  \textbf{10.608K}  & \textbf{0.948B}  &  60.1 & 69.49 & 72.35 & 54.56   \\ \cline{2-8} 
                   & SESR-M3~\cite{sesr} &  12.912K  & 1.154B  &   60.08 & 69.57 & 72.15 & 54.6   \\ \cline{2-8} 
                   & SESR-M5~\cite{sesr} &  17.520K  & 1.566B  &  \textbf{60.26} & 69.83 & 72.33 & 54.84  \\ \cline{2-8} 
                   & SESR-XL~\cite{sesr} &  113.3K  & 10.13B  &   60.16 & 69.47 & 72.35 & 55.04   \\ \hline
\end{tabular}
\end{center}
\end{table*}

\subsection{Robustness Results}\label{sec:expRob}
Table~\ref{tab:rob} shows the robustness of three classification networks (MobileNet-V2~\cite{mobilenetv2}, ResNet-50~\cite{resnet}, and Inception-V3~\cite{inceptionv3}) to various gray-box attacks when the images are upscaled using Nearest Neighbor interpolation and SR networks shown in Table~\ref{tab:PS}. Note that, before the images are upscaled (using either deep learning based SR or Nearest Neighbor), they pass through JPEG compression and wavelet denoising stages. From the results, three observations are clear:
\begin{enumerate}
    \item Tiny networks like FSRCNN~\cite{fsrcnn} and SESR~\cite{sesr} result in more or less the same amount of adversarial robustness as extremely large EDSR-base and EDSR~\cite{edsr} models.
    \item Against gray-box attacks, compact networks such as MobileNet-V2 are far less robust compared to larger CNNs like ResNet-50 and Inception-V3. For instance, the best accuracy achieved by MobileNet-V2 is about $44\%$ against the APGD attack, whereas ResNet-50 (Inception-V3) achieves up to $52\%$ ($72\%$) against the same attack.
    \item It is easy to think that if smaller SR networks perform very well against gray-box attacks, can we just use traditional upscaling instead of deep learning-based SR? The answer to this question is: \textit{No, we still need deep learning-based SR models to achieve some degree of defense}. This is particularly true for compact networks like MobileNet-V2 (and also for ResNet-50), where a large gap exists in robust accuracy between the Nearest Neighbor and SR models. Interestingly, Inception-V3 is a very robust CNN that recovers robust accuracy even with Nearest Neighbor upscaling (with JPEG + Wavelet).
\end{enumerate}

Hence, the proposed SESR-based adversarial defense is quite effective against gray-box attacks even though it requires significantly fewer computational and memory resources.

\subsection{Effect of JPEG}\label{sec:expJP}
Next, as a quick ablation study, we also analyze the impact of JPEG compression on adversarial robustness when it is used in conjunction with SR and wavelet denoising. As demonstrated in Table~\ref{tab:jp}, JPEG in conjunction with SR and wavelet denoising consistently outperforms just SR and wavelet denoising. Hence, JPEG provides an additional layer of robustness.
\begin{table}[]
\caption{Robustness Results: No-JPEG vs. JPEG}
\label{tab:jp}
\begin{center}
\scalebox{0.95}{
\begin{tabular}{|l|l|cc||cc|}
\hline
\multirow{2}{*}{Classification} & \multirow{2}{*}{SR} & \multicolumn{2}{c||}{No-JPEG}    & \multicolumn{2}{c|}{JPEG}     \\ \cline{3-6} 
                   &                    & \multicolumn{1}{c|}{PGD} & APGD & \multicolumn{1}{c|}{PGD} & APGD \\ \hline\hline
\multirow{5}{*}{ResNet-50~\cite{resnet}} & EDSR-base~\cite{edsr}                 & \multicolumn{1}{c|}{45.92}  & 48.15   & \multicolumn{1}{c|}{48.66}  & 50.56  \\ \cline{2-6} 
                   & EDSR~\cite{edsr}                  & \multicolumn{1}{c|}{46.67}  & 49.09  & \multicolumn{1}{c|}{46.43}  &  49.08  \\ \cline{2-6} 
                   & FSRCNN~\cite{fsrcnn}                  & \multicolumn{1}{c|}{46.71}  & 48.87  & \multicolumn{1}{c|}{49.8}  &  51.76 \\ \cline{2-6} 
                   & SESR-M2~\cite{sesr}                & \multicolumn{1}{c|}{44.94}  &  46.91 & \multicolumn{1}{c|}{49.66}  & 51.82  \\ \cline{2-6} 
                   & SESR-XL~\cite{sesr}                 & \multicolumn{1}{c|}{44.46}  & 46.04  & \multicolumn{1}{c|}{48.96}  & 51.24  \\ \hline\hline
\multirow{5}{*}{Inception-V3~\cite{inceptionv3}} & EDSR-base~\cite{edsr}                 & \multicolumn{1}{c|}{67.37}  & 67.39  & \multicolumn{1}{c|}{69.55}  & 72.17  \\ \cline{2-6} 
                   & EDSR~\cite{edsr}                  & \multicolumn{1}{c|}{67.43}  & 67.95  & \multicolumn{1}{c|}{69.57}  & 72.49  \\ \cline{2-6} 
                   & FSRCNN~\cite{fsrcnn}                  & \multicolumn{1}{c|}{66.39}  & 66.71  & \multicolumn{1}{c|}{69.93}  & 71.97  \\ \cline{2-6} 
                   & SESR-M2~\cite{sesr}                & \multicolumn{1}{c|}{66.81}  & 66.85  & \multicolumn{1}{c|}{69.49}  & 72.35  \\ \cline{2-6} 
                   & SESR-XL~\cite{sesr}                 & \multicolumn{1}{c|}{67.23}  & 67.27  & \multicolumn{1}{c|}{69.47}  & 72.35  \\ \hline
\end{tabular}
}
\end{center}
\end{table}

\subsection{Latency on Arm Ethos-U55 micro-NPU}\label{sec:expU55}
Finally, we quantify the concrete latency improvements obtained by using SESR for adversarial defense on a commercial micro-NPU called Arm Ethos-U55. Since Ethos-U55 is a very small NPU (0.5 TOP/s) that is meant to accompany highly resource constrained microcontrollers like Arm Cortex-M-based systems, we have used only the MobileNet-V2 network for classification. As mentioned earlier, in the defense setting considered, the MobileNet-V2 network does not take the traditional $224\times224$ image (which requires about 300M MACs), but instead takes an upscaled $598\times598$ image. The upscaled image increases the MAC count for MobileNet-V2 from 300M to nearly 2.1B MACs. However, from Table~\ref{tab:PS}, we can see that traditional SR methods like FSRCNN take 5.82B MACs to upscale an image from $299\times299$ to $598\times598$, which is significantly higher than even the enlarged MobileNet-V2. This further highlights the importance of SESR models that significantly reduce the compute cost over other SR networks.

Table~\ref{tab:u55} shows the inference latencies\footnote{The performance estimator for Arm Ethos-U55 micro-NPU is publicly available at \url{https://git.mlplatform.org/ml/ethos-u/ethos-u-vela.git/about/}.} of the enlarged MobileNet-V2 and various SR networks. Clearly, in end-to-end latency (i.e., combined classification + SR), SESR-M2 significantly outperforms FSRCNN by achieving nearly $3\times$ higher FPS and about the same robust accuracy. Hence, SESR networks enable a highly efficient defense against gray-box attacks on a tiny microcontroller-scale micro-NPU. 
\begin{table}[]
\caption{Latency on Arm Ethos-U55: Enlarged MobileNet-V2 + SR}
\label{tab:u55}
\begin{center}
\begin{tabular}{|l|c|c|c||c|}
\hline
SR Model  & Classification & SR & Total & FPS \\
& Latency (ms) & Latency (ms) & Latency (ms) &\\ \hline
FSRCNN~\cite{fsrcnn} &  \multirow{4}{*}{46.18}                 & 143.73 & 189.91 & 5.26 \\ \cline{1-1} \cline{3-5} 
                  SESR-M5~\cite{sesr} &                   & 26.76  & 72.94 & 13.70  \\ \cline{1-1} \cline{3-5} 
                  SESR-M3~\cite{sesr} &                   & 22.38 & 68.56 & 14.58  \\ \cline{1-1} \cline{3-5} 
                  SESR-M2~\cite{sesr} &                    & \textbf{20.19}  & \textbf{66.37} & \textbf{15.06} \\ \hline
\end{tabular}
\end{center}
\end{table}

\section{Open Challenges}\label{sec:dis}
Based on our analysis, we highlight the following open challenges in SR-based gray-box adversarial defense, particularly for constrained autonomous systems:
\begin{itemize}
    \item For gray-box attacks, compact models like MobileNet-V2 are still significantly less robust than other networks such as ResNet-50 and Inception-V3. Hence, better methods are needed to make the compact networks more robust.
    \item Can we build even smaller SR networks that result in higher robustness? Specifically, while robust accuracy stays more or less the same for most SR networks, it falls significantly for cheap interpolation techniques (especially for MobileNet-V2). Hence, further research is needed to understand at what limit the upscaling-based defenses fail and robustness starts to decrease significantly.
    \item Finally, certain classifiers seem to be more robust to any upscaling technique (like Nearest Neighbor with Inception-V3) than others. Therefore, more research is required to shed light on how SR and classification networks work together to achieve robustness. This could be useful, for example, in creating both efficient SR networks and efficient classifiers for constrained autonomous systems.
\end{itemize}
Exploring the above challenges in future research can drive significant progress in hardware-efficient SR-based defenses.

\section{Conclusion}\label{sec:conc}
In this paper, we have explored the computational efficiency aspect of SR-based adversarial defense mechanisms. Specifically, we have answered the question: Does robustness suffer if we use highly efficient SR networks in the inference pipeline? To this end, we have exploited SESR networks -- that achieve the state-of-the-art in efficient SR -- to defend various classifiers against adversarial attacks. Our detailed experiments have demonstrated that tiny SR networks like FSRCNN or SESR achieve as much robustness as extremely large CNNs like EDSR. We have also shown that SESR achieves up to $3\times$ improvement in end-to-end latency compared to FSRCNN when performing the SR-based adversarial defense on commercial micro-NPUs like the Arm Ethos-U55. Thus, SESR networks make SR-based defense more realistic on tiny micro-NPUs. Being able to defend inference pipelines in such highly resource-limited scenarios is extremely important for modern hardware-constrained autonomous systems.

\bibliographystyle{IEEEtran}
\bibliography{egbib}

\end{document}